# Effective Vortex Pinning in MgB$_2$ thin films

Y. Bugoslavsky, L.Cowey, T.J.Tate, G.K. Perkins, J. Moore, Z. Lockman, A.Berenov,
J. L. MacManus-Driscoll, A.D. Caplin and L. F. Cohen
*Centre for High Temperature Superconductivity, Imperial College of Science Technology and Medicine, Prince Consort Road, London SW7 2BZ, UK*

H Y Zhai, H M Christen, M P Paranthaman and D H Lowndes
*Oak Ridge National Laboratory, Oak Ridge, TN 37931-6056, USA*

M H Jo and M G Blamire
*Department of Materials Science, Cambridge University, Cambridge CB2 3QZ, UK*

We discuss pinning properties of MgB$_2$ thin films grown by pulsed-laser deposition (PLD) and by electron-beam (EB) evaporation. Two mechanisms are identified that contribute most effectively to the pinning of vortices in randomly oriented films. The EB process produces low defected crystallites with small grain size providing enhanced pinning at grain boundaries without degradation of $T_c$. The PLD process produces films with structural disorder on a scale less that the coherence length that further improves pinning, but also depresses $T_c$.

## 1 Introduction

An advantage of MgB$_2$ superconductors over Nb$_3$Sn and NbTi is its higher critical temperature of 39K combined with its relatively low raw materials costs. These factors may make MgB$_2$ a serious contender for use in cryogen free applications at 20K. Although cooling costs at 20K are substantially lower than 4K operation, the high cost of HTS conductors has so far prevented this market from being realised despite the existence of many working demonstrators. MgB$_2$ shows substantial promise in being able to bridge this economic gap.

One of the key properties that determines the prospects for large-scale applications is the magnitude of the critical current density $J_c$. Bulk or powder MgB$_2$ has rather high values of $J_c$ at zero field ($J_c$ in excess of $5 \cdot 10^5$ A/cm$^2$ at 20K), but exhibits a fast decay of $J_c$ in applied magnetic fields [1], which results in low values of the irreversibility field. These problems impose strong limitations on prospective use of MgB$_2$ e.g. for superconducting magnets. In addition the fast drop-off of the current with field is accompanied by an increasing rate of magnetic creep [1]; both effects stem from the lack of effective vortex pinning centres. Understanding the pinning mechanism and identifying the possibilities of pinning enhancement are therefore issues of critical importance. In very general terms, the possible mechanisms can be divided into two classes. The pinning can originate from the short-range disorder within grains [2], or otherwise from longer-range inhomogeneities associated with the polycrystalline structure and grain boundaries [3,4]. Understanding which of these scenarios can provide most efficient pinning will help define an optimal route of engineering the material either by varying the grain size, or by chemical doping, or by introduction of inclusions of suitable size and density within grains. The latter will eventually degrade the superconductivity and $T_c$ will drop. Earlier we have shown that pinning in bulk fragments can be significantly improved by introducing point-like defects by proton irradiation [5]. However, as yet it is unclear which route – intra- or inter granular pinning is the most effective. Thin film studies may guide the optimal approach and this is the aim of the current work.

## 2 Experimental Details

### 2.1 *Film Growth*

In this work we compare the pinning properties of MgB$_2$ thin films grown by two methods: pulsed laser deposition (PLD) and electron-beam (EB) evaporation. A series of films were deposited by PLD on unheated sapphire substrates from a stoichiometric target prepared from commercial MgB$_2$ powder. Approximately 500nm thick films were deposited at a pulse repetition rate of 10Hz at a pressure of 3 Pa in a 4%H$_2$ in Ar gas mixture. The as-grown films were not superconducting, so they have been post-annealed at 750$^0$ C within a sealed capsule containing both Mg powder and foil in the vicinity of the film. The annealing conditions were different for each film; the details have been published elsewhere [6]. The films had a reduced $T_c$, broad transitions and high resistivity (ρ) in the normal state. The values of the critical temperatures we use in this work were determined from the onset transition point in magnetisation vs temperature measurements. In

- 2 -

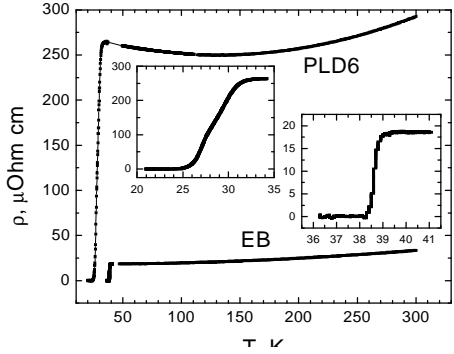

Fig. 1. Comparison of resistivity in two $MgB_2$ thin films. The PLD film appears to be more inhomogeneous, as indicated by depressed $T_c$, broad transition, high normal-state resistivity and activation-type dependence of $\rho(T)$ at low temperatures.

inhomogeneous PLD films the transitions were broad, and the "magnetic" transition was slightly lower than the onset of the resistive transition. For example, in the film PLD6 these points were at 29 K and 32 K, respectively (cf. Fig. 1). All the PLD films studied were randomly oriented, except one (PLD 8) that was crystallised at higher temperature of 950 °C and showed evidence of biaxial texture [6]. The EB film was manufactured at Oak Ridge National Laboratory [7] by first evaporating pure boron on a sapphire substrate, with subsequent reaction with Mg vapour. This film had a narrow transition at 38.7 K, and low dc normal state resistivity.

The $\rho(T)$ dependences of two films (EB and PLD6) are compared in Fig. 1. There may be up to 50% uncertainty in the absolute value of resistivity, due to the fact that the actual thickness of $MgB_2$ is not precisely known. Cross-sectional SEM [7], shows a 100 nm – 200 nm thick reaction layer next to the substrate in the EB film. The two films shown in Fig 1 have very different temperature dependences above $T_c$. The resistivity of the EB film has a metallic signature (i.e. $\rho$ decreases with decreasing temperature), whereas the opposite is the case with the PLD film below 120 K. The latter behaviour, together with this film's high resistivity indicates strongly non-metallic grain boundaries.

## 2.2 Film Microstructure and Electrical Connectivity

The EB film was deposited on A-plane sapphire (1120). The lattice mismatch inhibited epitaxial growth of the $MgB_2$. An XRD scan of the EB film showed sharp sapphire peaks and a broad, weak peak of MgO. No peaks belonging to $MgB_2$ were observed. The PLD film was grown on C-plane sapphire (0001) and the XRD showed the presence of highly crystalline Mg and poorly crystalline MgO. Again, no definitive $MgB_2$ peaks were observed. However the (101) $MgB_2$ peak at 42.4° may overlap with, and therefore be masked by the strong (006) sapphire peak at 41.7°. A pole figure scan of the (102) $MgB_2$ peak was then performed and this revealed no in-plane texturing. The absence of definitive $MgB_2$ peaks in both films may be due to poor crystallinity and/or a small volume of $MgB_2$, as well as lack of alignment. In addition, the weak scattering of the light elements contributes to the low diffraction intensities.

The surfaces of the films are shown in the SEM images presented in Fig. 2. The surface of the representative PLD film (PLD6) consisted of roughly spherical fragments about 1μm in diameter. The EB film appeared much better connected. More detailed microstructural studies have revealed that the size of the crystallites in polycrystalline thin films is of the order of 10 – 100 nm, with grain boundaries a few nm wide [4]. Hence we believe the 1-μm scale fragments seen in the SEM image are themselves made up of much smaller crystallites.

One of the key issues in determination of the critical current density ($J_c$) is the length-scale (L) over which continuous superconductive current paths are extended in the material. We have assessed the supercurrent connectivity using two kinds of measurements. The length-scale L of the supercurrent path can be estimated directly from magnetisation measurements [8]. Such an estimate (made at T = 10 K) showed that L was close to the films' lateral dimensions at zero applied field, with no significant decrease of L as the field increased to 3 Tesla. Fig 3 shows that the results were similar for both types of films studied, over a range of external fields. These results indicate that the grain boundaries act as effective superconducting contacts in both films and the ability to carry current across a grain boundary is not significantly degraded as a function of applied field. This conclusion was confirmed by direct mapping of the magnetic flux distribution in the EB film using a scanning Hall probe technique [9] as shown in figure 4(a). The flux density is directly related to the current distribution [10] and thus the magnetic imaging is a convenient tool for the current flow visualisation. The image was taken at 20 K and in zero magnetic field, after the film had been magnetised at 100 mT with the field applied perpendicular to the film plane. For a uniform film, a smooth roof-top profile would be expected, with straight bright ribs (current discontinuity lines) running along the central part of the film and diagonally from the corners. This pattern is visible in the left-hand side of the image. The right edge of the film is not straight, which smears the discontinuity lines. The presence of the roof-top structure is a direct evidence of the global film connectivity. The local distortions in the image can be understood by comparing the magnetic and optical image (see figure 4(b)). The film has a large number of almost circular holes up to 1 mm in diameter, through which the substrate can be seen. It appears that the corrugations in the magnetic image are consistent in density with the density of smaller voids, and the major black spots (indicating opposite direction of the magnetic induction, which is expected to be seen at the film edges) clearly correspond with the large holes at the left side of the film.

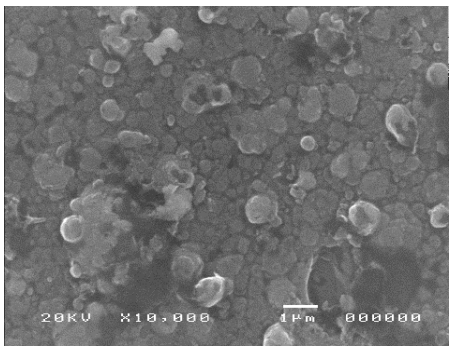 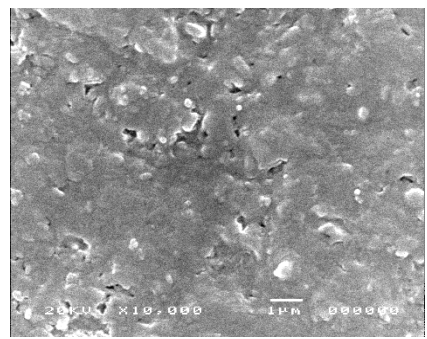

Fig. 2. SEM images of differently prepared films. Left: PLD6 film consists mainly of spherical grains about 1 μm in diameter. Right: In the EB film, the surface looks monolythic, although outgrowths and voids are also visible.

The conclusion from this study is that, in spite of granularity and large-scale defects in the films, they were continuous in the sense of global supercurrents and there was no evidence of weak-link behaviour. This fact validates the application of the standard procedure to extract $J_c$ from magnetisation data, and it confirms that the field dependence of the magnetic moment genuinely reflects the behaviour of $J_c$ rather than fragmentation of the current flow. However, the absolute values of the evaluated current density still remain a subject of uncertainty. From the SEM micrographs, the films are clearly less than 100% dense. The sparse granular structure (especially in the PLD films) is likely to impose significant geometrical restrictions on the current flow. The actual area of inter-grain contacts may well be smaller than the overall film cross-section. Therefore the inferred numerical values for $J_c$ are necessarily only a lower bound of the local intergranular current density. The degree of the underestimate is dependent upon the sample density, and therefore it ought to be more significant in the PLD films.

### 2.3 The critical current density and the irreversibility line

The critical current density was extracted from the width of the magnetic hysteresis loops that were obtained using a vibrating sample magnetometer. The irreversible magnetisation was converted into $J_c$ using the Bean model [11]. The validity of this procedure in the applied magnetic field was confirmed by the above connectivity analysis. The latter proves that the decreasing magnetisation in increasing field is due to a genuine decrease of $J_c$ rather than to the sample fragmenting into weakly-linked regions. Whereas we can be certain about the field dependence of $J_c$, its absolute value is to some extent imprecise, because the film thickness is known only within a factor of two.

Previously we reported a relationship between the high field behaviour of $J_c$ and the $T_c$ for a series of films [6]. However, we have since appreciated that magnetisation measurements of superconducting films are subject to strong systematic errors at high field due to the limited field homogeneity in our vibrating sample magnetometer. The measurement error becomes pronounced in the regime when the sample penetration field is comparable with the variation of the applied field along the travel of the vibrating sample, i.e. when the $J_c$ is small and H is high. We therefore consider the high-field data and, in particular, estimates of the irreversibility line as qualitative.

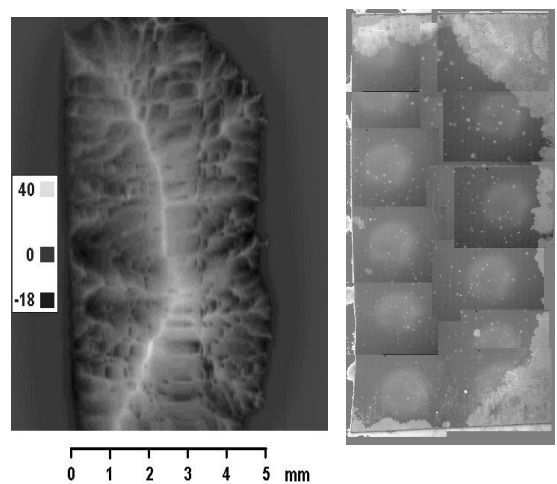

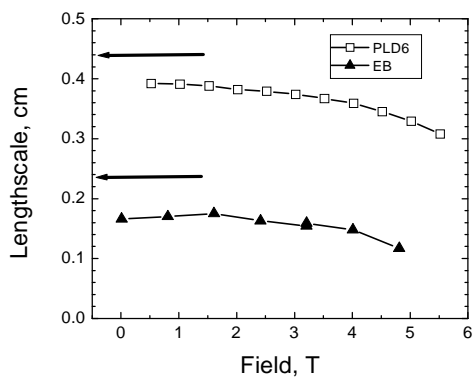

Fig. 3. The lengthscale analysis of the current flow in the films. As the field increases the lengthscale remains close to the lateral film dimensions, indicated by the horizontal arrows. The magnitude of the lengthscale and the fact that it is nearly constant with field confirms connectivity of the supercurrent.

Fig. 4. The scanning Hall probe image of the EB film (left) and the optical micrograph of the same sample (right). The colour scale is in units of mT. The optical image was compiled from separate fragments, hence its mosaic appearance. The Hall-probe mapping reveals a globally connected current flow; the fine structure in the magnetic induction distribution is likely to stem from numerous sub-millimetre voids in the film. These voids are clearly seen in the optical image as bright spots where light reflects from the polished substrate.

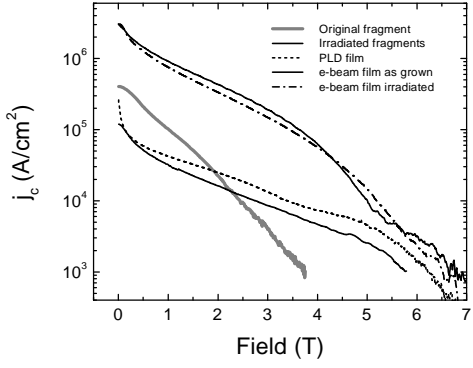

Fig. 5. $J_c$'s of various $MgB_2$ samples at 20 K.

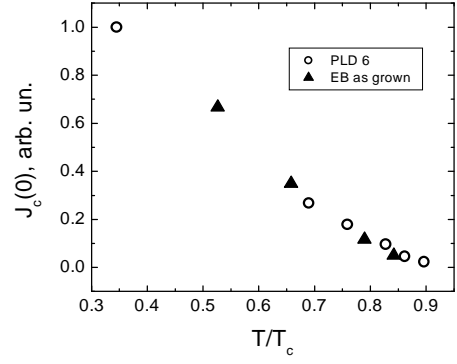

Fig 6 The zero-field $J_c$ normalised to its low-temperature value, as a function of reduced temperature, for PLD6 and EB film.

*2.4 Irradiation of films*

Films were irradiated using the Whickham implanter at Imperial College. $^1H_2$ molecules were implanted at an energy of 50keV, to a dose of $2.5 \times 10^{13}$ molecules cm$^{-2}$, at room temperature. This is equivalent to implanting $5 \times 10^{13}$ atoms cm$^{-2}$ of $^1H_1$ at 25keV. These parameters were chosen in order to localise the irradiation damage in the first 300nm of the film, modelled using SRIM [12]. The SRIM model predicts a roughly triangular damage profile, peaking at about 250nm, and with an average value of 0.04% displacement per atom (dpa) [13] over the thickness of 300nm. Two sequential irradiations were performed on the same sample, giving doses of $5 \times 10^{13}$ atoms cm$^{-2}$ (low dose) and $1 \times 10^{14}$ atoms cm$^{-2}$ (high dose), equivalent to dpa of 0.04% and 0.08%, respectively.

**3  Results**

For application of $MgB_2$, it is important to compare material performance at a fixed absolute temperature. Fig 5 shows the field dependence of $J_c$ at 20K for a representative PLD film (PLD6), the EB film and for comparison, the virgin and irradiated bulk fragments [5]. Over a reasonably wide field range the current has an exponential dependence on field (i.e. closely linear on the semi-log plot), $J_c \sim 10^{-H/H_{dec}}$. We consider the rate of the $J_c$ decrease in applied field to be directly related to the effectiveness of the pinning. As the thrust of this work is on the improvement of the high-field performance of $MgB_2$, the value of the decrement field, $H_{dec}$ can be regarded as a technical figure of merit. Higher values of $H_{dec}$ indicate more effective pinning properties. It is worth noting that, with the present lack of precision in the numerical value of $J_c$ on the one hand, and the good *relative* accuracy of the field dependence of $J_c$ on the other, the decrement field appears to be the only pinning-related parameter that can be considered without reservations.

As it can be seen in Fig 5, the irradiated fragment and both types of films have much more gradual field dependence of $J_c$ than the original fragment. The EB film is superior to the PLD film in the whole filed region. The difference in the absolute values of $J_c$ between the two films is significant (a factor of 30 at zero field) and probably can not be explained merely by the grain-contact area argument. The higher value of $J_c$ in the EB film is to be expected as it has a much higher $T_c$. Even comparing the $J_c$ values at the same reduced temperature, $T/T_c = 0.3$, we find that the EB film is superior at zero field, with $J_c(0) = 8.5 \cdot 10^6$ A/cm$^2$ whereas the PLD6 film has $J_c(0) = 1.5 \cdot 10^6$ A/cm$^2$. At the same time, the value of $H_{dec}$ (which is of course independent of the critical current calibration) is higher for the PLD film.

To analyse the temperature dependence of $J_c$, we have adapted the scaling methodology frequently used for HTS materials [14]. In Fig. 6 we plot $J_c$ (B=0, $T/T_c$) normalised by the low-temperature values of $J_c$. Temperature controls the $J_c(B=0)$ value in the same way in both films. The inaccuracy in our determination of $H_{irr}$ prevents us from performing a similar scaling of $J_c(B)$ curves. However, we have already established that at low fields the functional form of $J_c(B)$ is similar for all the films studied. It is interesting that, although the effectiveness of the pinning in absolute terms is different, the zero-field $J_c$ varies with temperature in a linear manner in both films.

Interestingly, the linear temperature dependence is observed for $H_{dec}$ in both kinds of films as well (Fig.7), albeit with very different gradients. These data suggest that in the PLD film the pinning efficiency increases with decreasing temperature at a much higher rate than in the EB film.

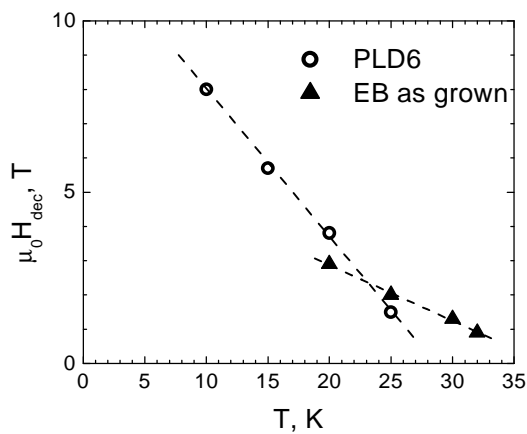

Fig 7. The decrement field as a function of temperature PLD6 and EB films.

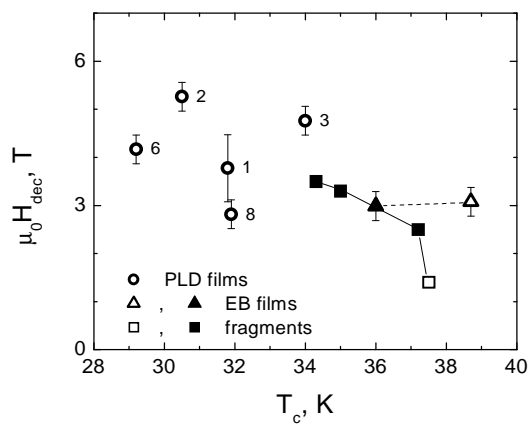

Fig 8: The decrement field $H_{dec}$ as determined at T=20 K and H=2 T. $H_{dec}$ is a characteristic of the rate with which the current decays as the field is increased. The data for original and proton-irradiated fragments, the as-grown and irradiated EB films, and different PLD films are presented, all measured at 20 K. The fragments were irradiated to nominal damage levels of 0.04%, 1% and 5%, the increase in dose corresponding to the decrease in $T_c$. The irradiation of the EB film resulted in reducing $T_c$ but no significant variation of $H_{dec}$. The PLD films have reduced $T_c$ and high values of $H_{dec}$.

In order to clarify the explicit dependence of the pinning effectiveness on $T_c$, figure 8 shows a plot of $H_{dec}$ (as determined in the interval between 1 and 2T) for a series of PLD grown films with varying $T_c$, plus the as-grown and then irradiated EB film. The data for the non-modified and the proton-irradiated fragments [5] are also shown for comparison. In the polycrystalline fragments, the increasing irradiation dose enhances $H_{dec}$ while bringing the $T_c$ down. There is no such an exact correlation between $T_c$ and $H_{dec}$ in the PLD films. The large scatter of the data for different PLD films highlights the fact that the $T_c$ is far from being the only parameter that reflects the pinning properties. Other factors such as partial grain texture may well play a role. (Indeed, the film PLD8 was partially biaxially textured [6], and its pinning properties are less effective than in PLD1- a film with the same $T_c$ but with randomly oriented grains.) However the population of PLD films occupies an area on the plot that can be seen as a continuation of the trend outlined by the irradiated fragment data. The as grown EB film stands well clear of this trend. Not only has it relatively high $H_{dec}$, but its $T_c$ is even higher than that of the original fragment. The estimated zero-field $J_c$ of this film is also high. This behaviour indicates that another kind of pinning disorder is possible, which comes at no expense to $T_c$. Still as far as $H_{dec}$ is concerned, the PLD films perform noticeably better. This is valid when the comparison is made at the same absolute temperature of 20K, which corresponds to different reduced temperatures. The $H_{dec}$ value for the PLD6 film at 15 K (so that $T/T_c \approx 0.5$) is 6T (see Fig. 7), twice as high as that of the EB film at 20 K and the same $T/T_c$. Quite naturally, these results put forward the question whether the pinning properties EB film can be further improved (i.e. $H_{dec}$ increased towards the values demonstrated in the PLD films) while possibly retaining good $T_c$ and $J_c(0)$.

The proton irradiation of the EB film was performed in order to check this possibility. From the SRIM calculation, the proton beam energy and dose should have produced a nominal defect density only a tenth of that used in the fragment study [5]. However the thickness of the fragments was much bigger than the film thickness, so the beam energy was a factor of 10 higher in the former case. The damage profile through the sample thickness is therefore likely to be different, and it is not surprising that there is a discrepancy in the apparent efficiency of the irradiation in films and fragments. The modest dose modified the EB film significantly and $T_c$ dropped by 3K (see figure 8) as a result of two irradiation steps. However, from figure 5 it can be seen that almost no changes were made in the pinning properties. The irradiation may have slightly improved the performance at high field but the difference is within the experimental error.

This result, although negative in terms of improving the pinning, may prove useful in understanding possible origin of defects in the as-grown film. From the drop in $T_c$ we know that the irradiation has produced a sizeable number of defects in the film, and the distance between the defects is comparable to the coherence length. In this respect, the effect of irradiation is very similar in the films and the fragments, and it can be assumed that the nature of defects is similar in the two experiments. Then the lack of pinning enhancement is an evidence of a different kind of disorder in the as-grown film. This disorder provides stronger pinning than the irradiation-induced defects might have created, therefore the effect of the latter is just not observable.

## 4 Discussion and conclusions

Crystallographic defects are present in the films both inside the grains and at the grain boundaries. The question is whether the intragranular or intergranular regions dominate the pinning properties. We assume that $T_c$ is lowered below 39K because of structural disorder on a length scale less than the coherence length, which is likely to be of the order of 3-5 nm in these films. The exact nature of the disorder leading to depressed $T_c$ is not yet known and is a topic of speculation. Nevertheless we can state that the greater the intragranular disorder, the lower the $T_c$.

Randomly oriented polycrystalline thin films may be very different to well-aligned epitaxial films. In particular, a correlation between pinning and dc resistivity may exist in the latter case [2] but not necessarily in the former. In randomly oriented films, there are two types of geometric constraints on the cross-section for current flow: polycrystalline grain boundaries approximately 3 nm wide between grains of dimension 10-100 nm (i.e. perfectly sized to be efficient pinning sites) [3], and boundaries between fragments with dimensions of the order of 1 μm. The fragment boundaries that are visible in SEM images will dominate the electrical connectivity and therefore the normal-state dc resistivity, but would not significantly contribute to pinning efficiency.

The EB film has much stronger pinning than the original bulk fragment, although both have $T_c$ values close to 39K. The electron-beam growth technique allows $MgB_2$ to form under equilibrium conditions producing bulk-like $T_c$ value and the high $T_c$ indicates a good degree of local order. We must therefore conclude that in the EB film defects are present that are at the same time strong and spaced at distances larger than ξ. In the context of the $MgB_2$ research, it is plausible that these strong pinning sites originate from the grain boundaries. The SEM image (figure 2b) supports this conclusion because the film has the appearance of a sintered polycrystalline ceramic, with very small grains. The irradiation study provides additional conformation because the pinning properties of the film were unaffected although $T_c$ was depressed after irradiation.

Grain boundary pinning must play a similar role in the PLD films. Indeed, the higher $T_c$ PLD films have values of $H_{dec}$ similar to the EB film. However, the PLD films are formed by ex-situ annealing of species which are mixed on an atomic scale. The large heat of formation of $MgB_2$ means that a metastable state could readily be formed (where impurity atoms such as O and C are retained in the crystal structure). As a result it is likely that these films, in general, are more disordered intragranularly. We have no evidence to claim that the structural defects in the PLD films and the irradiation-induced damage have similar *microscopic* nature, nevertheless, on the phenomenological level, they do cause similar pinning properties in the PLD films and irradiated fragments.

In respect to the high-field behaviour of $J_c$, the low $T_c$ PLD films are superior to the higher-$T_c$ films when compared at the same absolute temperature of 20 K. The comparison is further in favour of the PLD films, when made at the same reduced temperature. Grain boundaries are most probable to be heavily disordered in all PLD films and it is unlikely that they will be more defective in lower $T_c$ films.

The behaviour of PLD8 is somewhat anomalous because it has a reduced $T_c$ value but this appears to have done little to improve its pinning properties (over that of the EB film). However, this film was crystallised at a significantly higher temperature than the rest of the series so it is likely to have larger grains. The interplay between grain size (larger grain size probably reduces pinning), grain orientation (texture probably reduces pinning) and reduced $T_c$ value (intragranular disorder increases pinning) could lead to an overall decrease in pinning.

In summary, the comparative study of the pinning properties of differently prepared $MgB_2$ thin films reveals that both inter- and intra-granular pinning mechanisms are present. The EB process produces films with small grains that provide efficient pinning without depressing the $T_c$. This mechanism is of the inter-grain nature and most likely related to grain boundaries. In addition, improvement in pinning can be produced by progressively disordering the material intragranularly, on a lengthscale smaller than the coherence length, as it is the case with PLD films and proton-irradiated fragments.

Grain boundary orientation and intragranular disorder hold the key to the improvement of pinning in thin films. Well ordered, large grained, textured films will pin vortices less effectively but may offer other advantages, such as providing a better template for the reliable patterning of weak links for electronic applications.

## Acknowledgement

This work was supported by the UK Engineering and Physical Sciences Research Council.

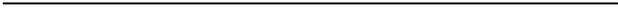